\documentclass[11pt,twoside]{article}
\usepackage{newpasp,epsf}
\def\edcomment#1{\iffalse\marginpar{\raggedright\sl#1\/}\else\relax\fi}
\reversemarginpar

\begin{document}
\title{3D MHD Simulations of Radio Galaxies Including Nonthermal Electron Transport}
\author{T. W. Jones \& I. L. Tregillis}
\affil{School of Physics and Astronomy, University of Minnesota,
Minneapolis, MN, 55455, U.S.A.}
\author{Dongsu Ryu}
\affil{Department of Astronomy \& Space Science, Chungnam National
University, Daejun, 305-764, Korea}

\begin{abstract}
We report on an effort to study the connections between dynamics in
simulated radio galaxy plasma flows and the properties of nonthermal
electron populations carried in those flows. To do this we have
introduced a new numerical scheme for electron transport that
allows a much more detailed look at this problem than has been possible
before. Especially when the dynamics is fully three dimensional
the flows are generally chaotic in the cocoon, and the jet itself
can flail about violently. The bending jet can pinch itself off
and redirect itself to enhance its penetration of the ambient medium. 
These behaviors often eliminate the
presence of a strong jet termination shock, which is assumed 
present in all modern cartoon models of the RG phenomenon. Instead a much more
complex ``shock web'' forms near the end of the jet that leads to a far
less predictable pattern of particle acceleration. Similarly,
the magnetic fields in these flows are highly filamented, as well as 
spatially and temporally intermittent. This leads to a very
localized and complex pattern of synchrotron aging for relativistic electron
populations,
which makes it difficult to use properties of the electron
spectrum to infer the local rate of aging.
\end{abstract}

\section{Introduction}
The interaction between high power plasma jets and the circumgalactic
medium (CM)  is now the firmly established paradigm for radio galaxies (RGs), 
as many talks at this meeting verify. The original ``twin exhaust'' model
(Blandford \& Rees 1974) outlined how fast plasma jets could possibly
carry energy efficiently from the active galactic nucleus (whose nature
was still the subject of speculation in 1974) into hot spots of 
radio lobes, depositing that energy at ``working surfaces'' associated
with the regions where the jets impinged on the ambient medium.  Blandford \& Rees
touched on most of the issues that have occupied researchers
in this field over the quarter century since. Those authors perceptively
captured a basic concept that seems now clearly to be at the heart of
the physics of the radio galaxy phenomenon, and also that previewed
discovery of remarkably similar outflows from an amazing diversity of
astrophysical systems.
We are still at the task of learning about jet physics more than 25 
years later, because 
the application of the simple concept is actually very complex. 
Jet-based flows are highly driven systems, inherently
unstable and probably always far from any general equilibrium. We
still do not properly understand how the jets are formed, what they are
made of, how they manage to
propagate as much as megaparsecs into the ambient medium, nor how
and where the relativistic electron populations we observe in the flows are
accelerated nor how they evolve more generally. Those issues, of course,
are the central themes of this meeting.

Over the past 15 years or so numerical simulations of time dependent
jet flows have
progressed enormously from the earliest two dimensional
axisymmetric gasdynamical
flows (Smith et al. 1985) to three dimensional flows
(e.g., Cox, Gull \& Scheuer 1991)
until now fully three dimensional
flows incorporating self-consistent MHD are relatively straightforward, 
if not yet easy, to model with modest resolution
(e.g., Clarke 1997). These simulation methods have also been extended
to include flows in either 2D or 3D with relativistic bulk motions 
(e.g., van Putten 1993;
Duncan \& Hughes 1994; Aloy et al. 1999; Zhang, Koide \& Sakai 1999).
Simultaneously, physical models of particle acceleration physics, especially as it
relates to the formation of collisionless shocks, have become
much better developed (e.g., Jones 2001), even if that cannot yet be
called a solved problem. All of these very positive developments have been
well represented in presentations at this meeting, in fact.

Most of our information about RGs currently derives from radio synchrotron
emissions reflecting the spatial and energy distributions of relativistic
electrons convolved with the spatial distribution of magnetic fields.
X-ray observations, especially of nonthermal Compton emissions,
depending on the electron and ambient photon distributions,
are now beginning to add crucial information, as well.
Using these connections, much effort has been devoted to interpreting 
observed brightness, spectral and polarization properties of the nonthermal
emissions for estimates of the key physical source properties, such as the
energy and pressure distributions and kinetic power, as well as to
find self-consistent models for the particle acceleration and
flow patterns. As telescopes and analysis techniques have improved
the level of detail obtained, it has become apparent, however,
that the observed properties are not very simple and much harder
to interpret than most simple models predict (e.g., Rudnick 2001).

This problem stems in part from the complexity of the flow dynamics
expected, and also for inherent difficulties in simulating both the
dynamics and the transport of radiating particles in multi-dimensional
flows. The latter is especially challenging. 
Until now, in fact, all published efforts to model nonthermal
emission properties from simulated 
flow behaviors have been based on ad-hoc simple assumptions about
the relationships between nonthermal particles and bulk flow variables 
such as total fluid
pressure, density and magnetic field (e.g., Clarke, Burns \& Norman 1989;
Matthews \& Scheuer 1990; Aloy et al. 2000). To address properly
the inherently nonequilibrium character of nonthermal particles,
however, it is essential that they be treated explicitly. 
We report here
on our program to do this. It provides the first multi-dimensional
numerical simulations of jet-driven flows including acceleration and
transport of radiating nonthermal electrons in a fashion that enables
detailed study of the connections between flow
dynamics, particle transport and nonthermal emissions. This is
made possible by a new and very efficient scheme for nonthermal 
particle transport (Jones, Ryu \& Engel 1999; Jun \& Jones 1999). We focus in
this paper on the links between plasma dynamics and particle
acceleration and transport that we can see with this treatment. 
A companion paper (Tregillis, Jones \& Ryu 2001a) discusses
initial ``synthetic radio observations'' made on our simulated
objects. 
We will see, in fact, that the nonthermal
particle and emission properties are often very poorly represented
in standard cartoons of RGs.

\section{Computational Methods}
Flow dynamics in our simulations is treated with a second-order, conservative
``TVD'' ideal MHD code described in Ryu \& Jones 1995, Ryu, Jones \& Frank 1995 and
Ryu et al. 1998. The method depends on approximate solutions to the
1D MHD Riemann problem at zone boundaries, and uses conventional
directional splitting techniques in multiple dimensions to retain
second order accuracy. The code maintains the divergence free condition for magnetic
fields to machine accuracy using an upwinded constrained transport
scheme as described in Ryu et al. 1998. To follow nonthermal particle
transport we solve the standard ``convection diffusion'' equation
for the momentum distribution, $f(p)$ (e.g., Skilling 1975). This
equation includes the effects of adiabatic and radiative losses
as well as terms that account for particle acceleration due to 
spatial diffusion at shocks and momentum diffusion resulting from
MHD turbulence. The computational effort needed to solve this equation
over an entire grid is enormous, because it must be solved simultaneously
for momenta spanning at least several orders of magnitude, and must
capture microphysics taking place on spatial scales also spanning
many orders of magnitude. To manage that with conventional finite
differencing methods in momentum space would involve vastly more effort than the
MHD itself, so is simply not practical. On the other hand, there are
a couple of very important properties of the problem that can be utilized
to circumvent the difficulties in some circumstances, reducing the work level 
to being just comparable to the MHD; that is, the
total cost and number of variables is about doubled, so manageable. 

The first key feature is that $f(p)$ is mostly a smooth
and broad function. In fact, away from cutoffs it can be adequately
described by defining a ``local spectral index'', $q(p) = - \partial \ln{f}/\partial \ln{p}$,
which varies slowly with $p$. 
Thus, we integrate the convection diffusion equation over logarithmic
momentum bins at each spatial zone and simply track the number of nonthermal particles in each
bin. Those quantities are updated in a conservative scheme
using fluxes across momentum boundaries computed from the dynamical variables,
magnetic field, radiation field, etc. By applying the simple quasi-power law
model for the distribution of $f(p)$ within momentum bins we can
make the bins rather large and still maintain good accuracy.
For the results shown here we have used eight momentum bins uniformly spread over
$\Delta \ln(p) = 12$.

The second key feature is that the microphysics
governing $f(p)$ generally takes place on scales within a modest
factor of the gyro radii of the particles. For energies of
a few tens of GeV and below, relevant to radio synchrotron emission and
X-ray Compton emission those length scales and the associated
time scales are at least several orders
of magnitude smaller than scales resolved in the MHD solution. 
Diffusive acceleration up to GeV energies at shocks should be essentially instantaneous 
on the time scale of RG dynamics, so 
one can find $f(p)$ immediately
behind a shock from the steady state solution there. In the test particle
limit that depends only on the shock compression ratio, r, i.e.,
$q_s = 3r/(r - 1)$, which  approaches 4 at strong shocks. Downstream,
$f(p)$ becomes modified by adiabatic and radiative
effects using the methods outlined in the previous paragraph. This scheme
can be implemented to include spatial diffusion, and second order Fermi
acceleration, as well as other loss mechanisms. In the present, exploratory
simulations, however,  we have neglected spatial diffusion, and have
included only diffusive shock acceleration, plus adiabatic and radiative losses for
nonthermal electrons. We treat the nonthermal electrons as a passive 
component for now, and have included improvements that enhance
the performance of the originally published scheme.

\section {Model Dynamics}
Jones et al. 1999 carried out several exploratory simulations using these
methods for axisymmetric jets. Their purpose, as for us also, was to
begin an examination of how best to understand the links between
complex RG flow dynamics and nonthermal particle transport properties.
They discussed three simulations based on identical flow dynamics, but
with different simple idealized models for the nonthermal particle
transport properties. Here we discuss extensions of those same three
simulations from two to three dimensions. Earlier discussions of
some of these results are contained in Jones, Ryu \& Tregillis 2000 and Tregillis, Jones \& Ryu 2000.

The 3D jets were light ($\rho_j/\rho_a = 10^{-2}$) with an internal
Mach number, $M_j = 8$, and assumed to be in pressure balance with
a uniform ambient medium (the CM) at their origin. 
The magnetic field in the in-flowing jet
was helical, with a uniform poloidal component, $B_{p0}$, that mapped into the CM,
plus a toroidal component, $B_t$, derived from a uniform axial current with a
return current along the jet boundary. The maximum $B_t/B_{p0} = 2$, while
the axial ``beta'' of the jet plasma was $\beta_0 = 8\pi P_g/B_{p0}^2 = 10^2$.
To break axisymmetry the in-flowing jet was made to precess on a
5 degree cone with a period allowing approximately $5 \frac{1}{2}$
rotations during the simulation. Thus, our flows have a fully
three dimensional flavor very early on. Previous 3D simulations have
shown that even numerical perturbations will eventually cause jets to
deviate substantially from quasi-2D symmetry (Norman 1996), but only
after the jet has propagated many jet radii. On the other hand there
is good evidence that at least some jets really do precess 
(e.g., Condon \& Mitchell 1984; Mantovani et al. 1999; Sudou \& Taniguchi 2000).
The simulations were carried out on a $576\times 192\times 192$ grid 
with open boundary conditions except for the jet origin.
The in-flowing jet had a top hat velocity profile inside a radius
of 15 zones, with a thin sheath around it. Thus, the long dimension of
the computational domain was approximately 38 jet radii. These flows
are, therefore, still relatively ``young'' compared to most RGs. Our
initial objectives depend on maintaining a reasonably fine resolution of
the dynamical structures in the jets and their ``heads'', so the
large effort needed in these simulations constrains us to look at young
flows for the moment.

\begin{figure}
\plotfiddle{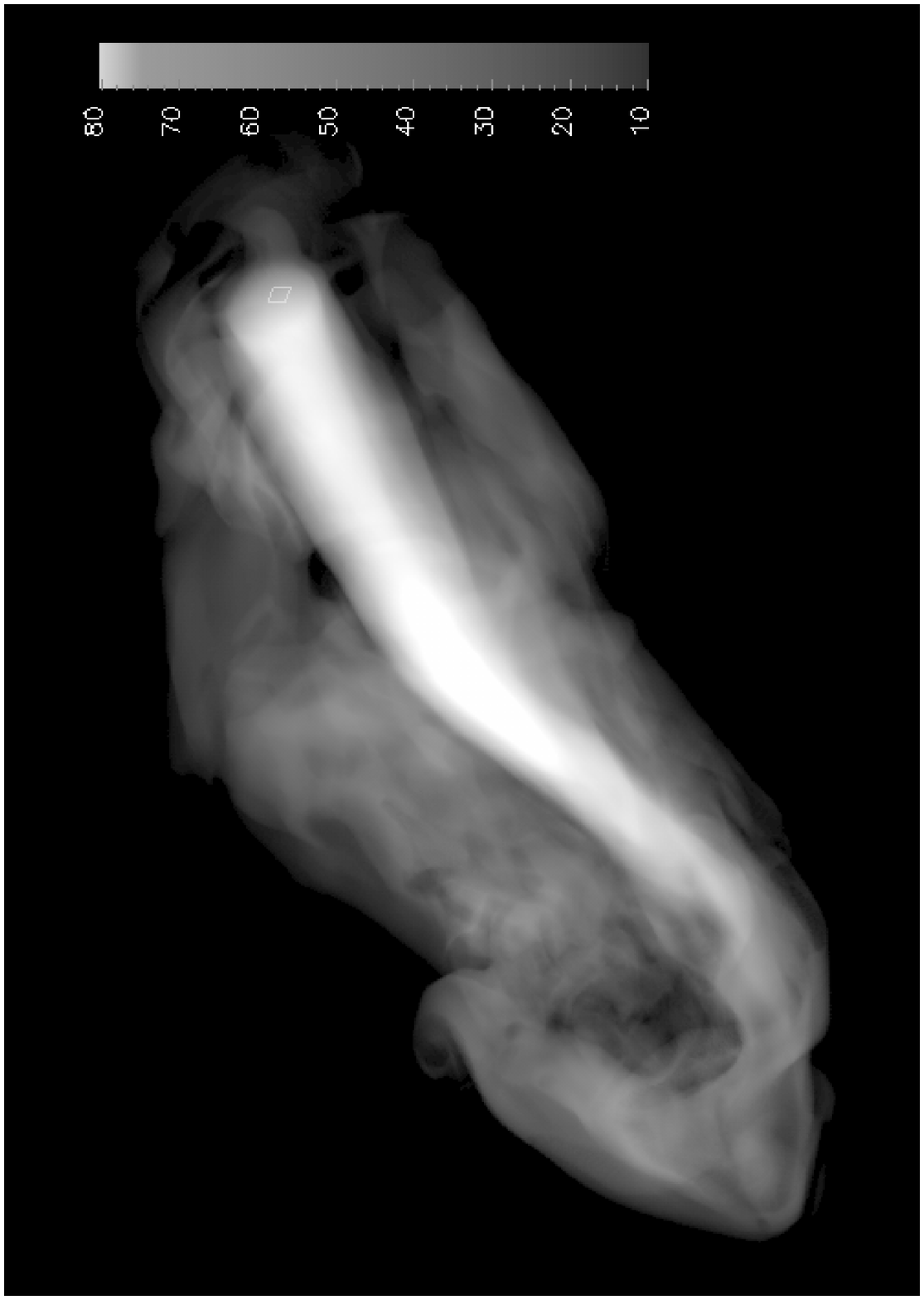}{1.7in}{-00}{30}{30}{-180}{-45}
\plotfiddle{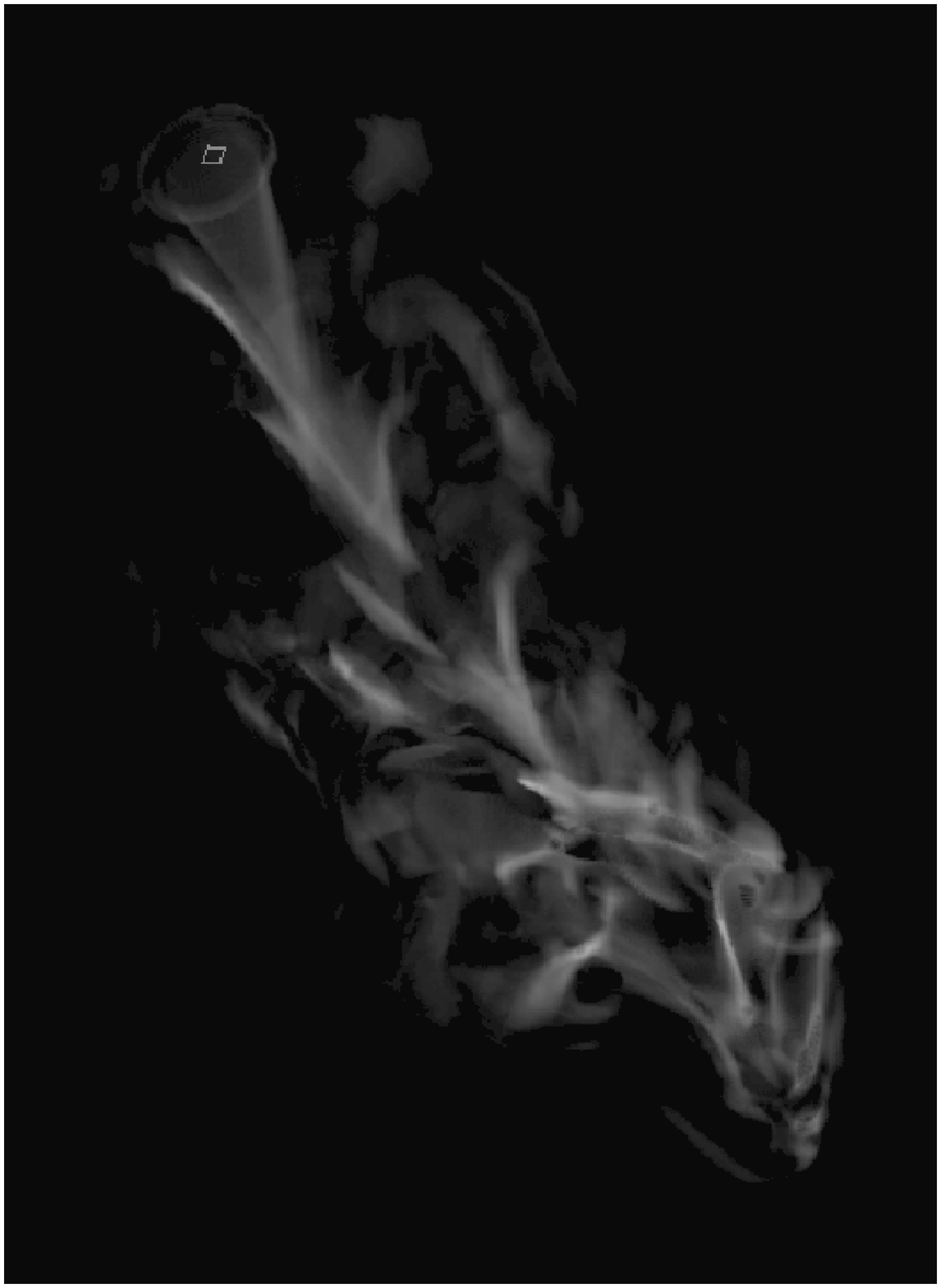}{0.0in}{-00}{30}{30}{-10}{-21}
\caption{Left: Volume rendering of the flow speed in a Mach 8 MHD
jet and its cocoon. The jet has penetrated the CM a little over
30 jet radii at this time. Only plasma entering the grid through the jet is
rendered visible here and in all the other figures, as well. Right: Shock
structures accompanying the flow shown.}
\label{fig1}
\end{figure}

\begin{figure}
\plotfiddle{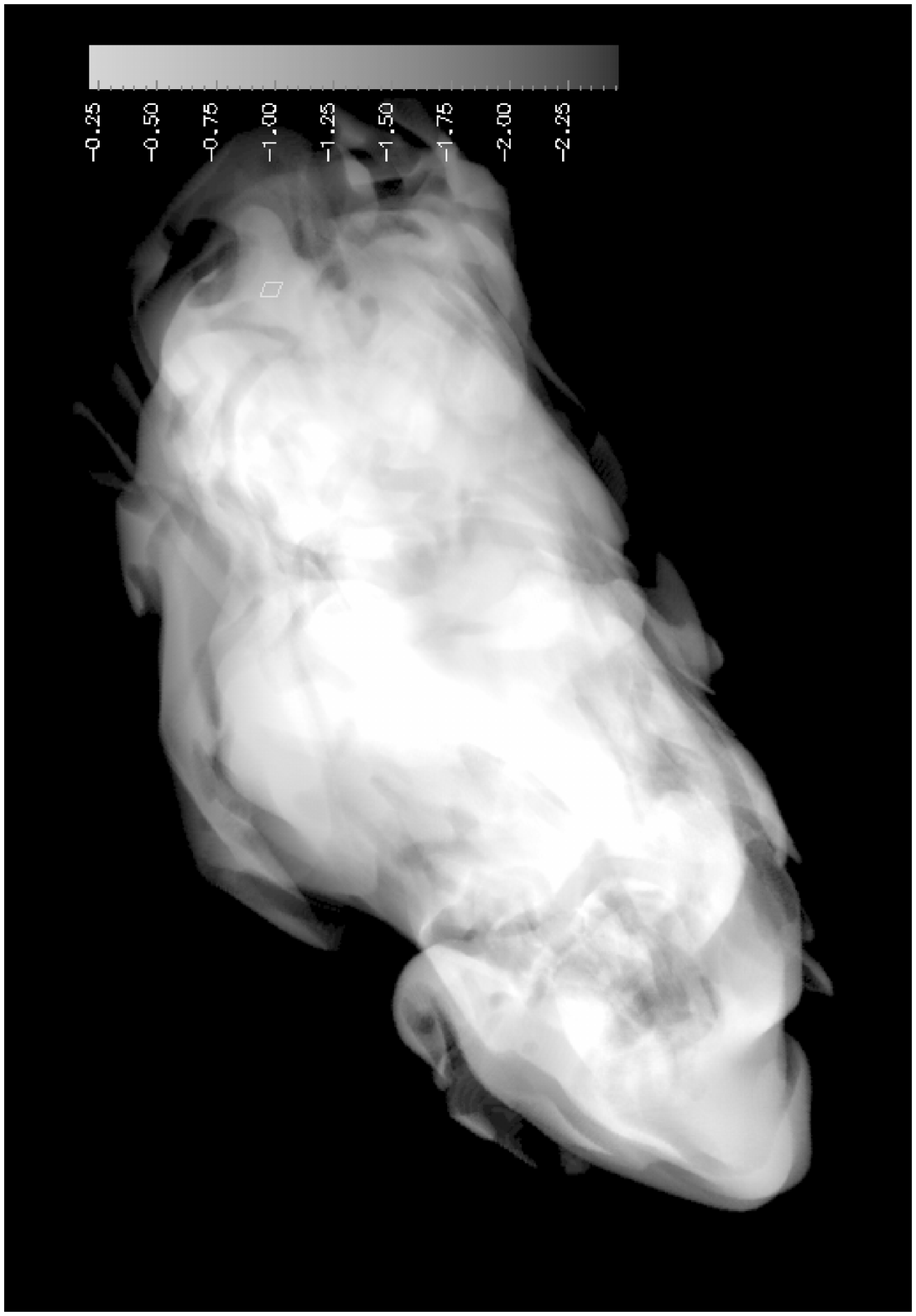}{1.5in}{-00}{30}{30}{-180}{-40}
\plotfiddle{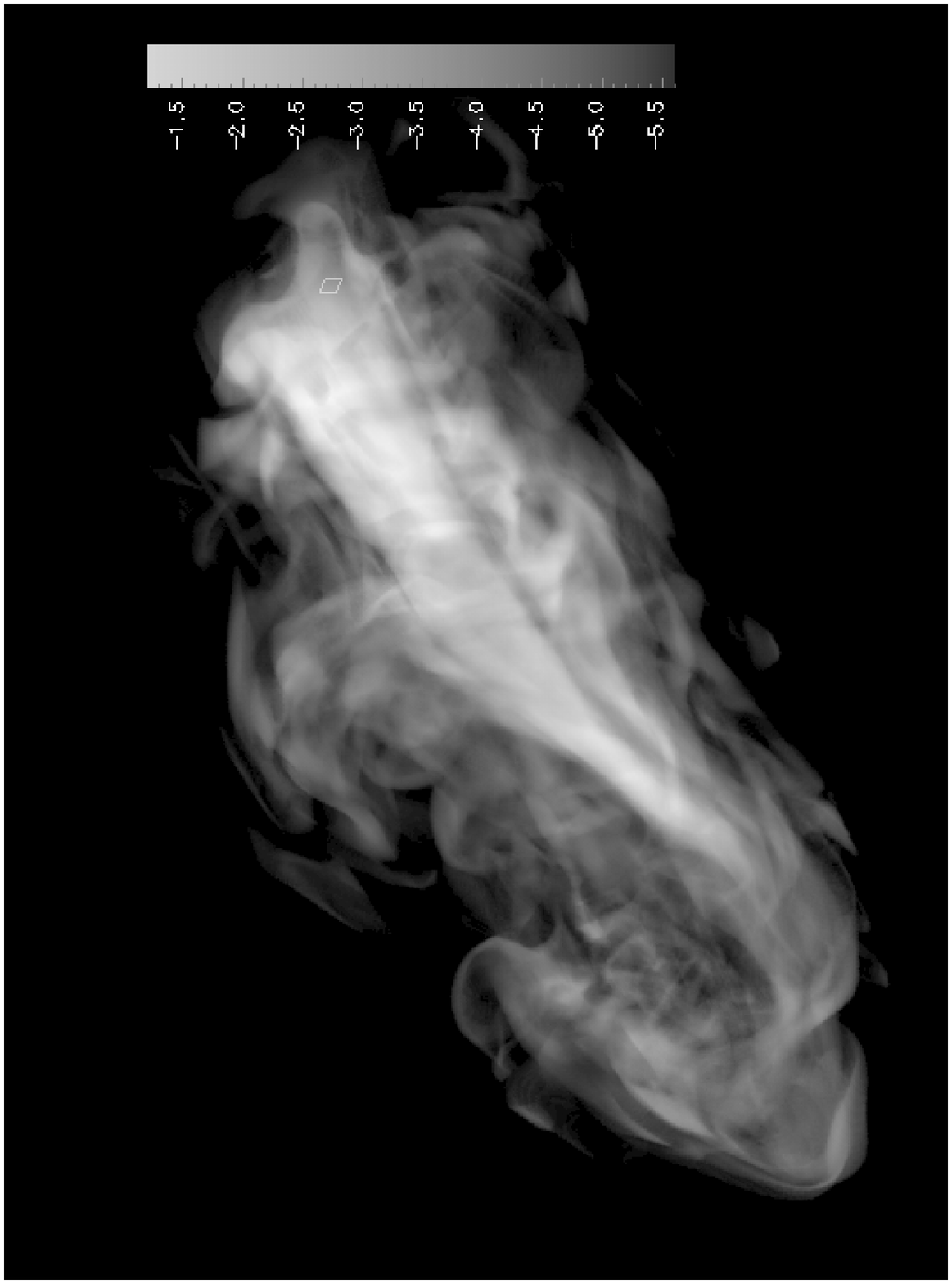}{0.0in}{-00}{30}{30}{-10}{-15}
\caption{Left: Volume rendering of the log of the gas pressure. Right:
The distribution of log of the magnetic pressure.}
\label{fig2}
\end{figure}

Figures 1 and 2 illustrate some of the basic dynamical properties of
the simulated flows about 90\% of the way to the end time. All of the
images shown here are volume renderings filtered to show only plasma
that entered the grid through the jet origin. That is, we have used
a passive mass fraction variable and made transparent any zone containing
less than 99\% jet material by mass.
Figure 1 represents the velocity field in this manner. The left panel displays
the flow speed of the jet and its cocoon. The jet itself is clear, of course,
but note also that flow speeds in the cocoon can be a substantial
fraction of the jet speed. It is also obvious that the cocoon flow is not
regular at all. That point is made more dramatic in the right panel
of Figure 1, which displays flow compression; i.e., $\nabla\cdot u$.
The image isolates shocks in the flow. While such classic features
as conical shocks near the jet origin are evident, it is hard to 
identify anything resembling the canonical ``jet termination shock''.
Rather, the head of the flow is filled with a ``shock web'', and most of
it involves back flows in the cocoon rather than the jet itself. There is
a very small jet termination shock near
the bottom right of the image, but close examination shows
that very little of the jet flow actually passes through it.

We have produced several animations, including flow speed, 
flow compression and magnetic pressure for the entire simulation in order
to study the flow properties and to explain the features seen here.
Those are currently available for viewing at URL\\
{http://www.msi.umn.edu:80/Projects/twj/radjet/radjet.html.}
Almost as soon as the jet terminus propagates past the
position of the first conical shock all semblance of any simple
symmetry vanishes. The end of the jet begins to ``flail'' violently,
occasionally isolating pieces of initially high speed plasma
that dissipate or run into the wall of the cocoon. The main jet
flow then extends forward abruptly, sometimes seeming
to poke a sharp ``finger'' into the CM. These structures resemble
hybrids of the so-called ``dentist's drill'' (Scheuer 1982) and
``splatter spot'' (Lonsdale \& Barthel 1986) concepts for
production of secondary hotspots, and confirm behaviors seen in some earlier
3D simulations (Cox, Gull \& Scheuer 1991). Most of the time there is
no recognizable jet termination shock, but the violence of the
end of the jet flow maintains a complex shock web similar to that visible
in Figure 1. Most of the shock surfaces are relatively weak compared to
what one would derive for the 1D jet termination shock (the latter strength
being for a light jet roughly the jet internal Mach number). 
Some portions of the shock web can, however, equal or exceed that strength.
Generally speaking flows within the cocoon in this simulation are
backwards directed, but also highly chaotic, as the shock web indicates.

Figure 2 illustrates the distribution of the log of gas pressure, $P_g$ and
magnetic pressure, $P_b$, for the same time as in Figure 1.
Note first that there is considerable intermittency to both
distributions. For $P_g$ this is reflects the complex flows within
the cocoon, since cocoon shocks will enhance $P_g$ and pressure
gradients drive the chaotic motions, as well. There is very little
similarity in the details of the distributions of $P_g$ and $P_b$, however.
The strongest magnetic fields are generated primarily by flow shear,
not compression, so this is expected. The filamentary nature of
the magnetic field is evident in the image, and it is clear that
the magnetic pressure is much more intermittent than the gas pressure,
as we would expect in a system that is this strongly driven and so
far from equilibrium. Local variations in $P_b$ of
two orders of magnitude are typical, with much larger excursions
into occasional magnetic voids. Peak field strengths in the cocoon are
comparable to the jet field, but the average cocoon field generally
is much less
on account of rarefaction in flows emerging from the jet.  
We cannot say, of course, how the magnetic
field would evolve on scales smaller than our numerical resolution, but
on our grid the magnetic field never approaches equipartition with the
thermal plasma; i.e., $\beta >> 1$ everywhere. There are times and
places where $M_A = (4\pi u\sqrt{\rho})/|B| < 1$, however, so that
magnetic stresses exceed Reynolds stresses. The magnetic field is
not entirely passive, in other words, even though on average it
would appear so.
We note also that the animation of the magnetic pressure reveals
brief episodes when the peak field strength near the jet terminus is considerably stronger
than average. Those episodes correspond to times when the
jet extends itself most rapidly into the CM as it ``breaks'' and
reforms.

In summary, the terminus of our light 3D jet behaves in a violently unstable
manner that makes the concept of a simple jet termination shock 
not very applicable, while creating a chaotic cocoon with a rich
web of shocks and highly filamentary, intermittent magnetic fields.

\section{Nonthermal Electron Acceleration and Transport}
On top of the dynamics just described we computed the evolution of
nonthermal electron populations using the methods outlined in \S 2.
We emphasize that our purpose at this stage of the program is not to
look for the parameters that necessarily most resemble real RGs, but, rather to
understand how particle populations will behave under simple assumptions
and how ``synthetic observations'' of the simulated objects 
behave. Further, we want to know and what that tells us that we can reliably
derive from observations when we know the actual source physical
properties.

We simulated three models for nonthermal electron transport analogous
to those discussed by Jones et al. 1999 for similar, but axisymmetric flows.
In all three models nonthermal electrons passing through shocks are
accelerated according to  standard test particle
diffusive shock acceleration theory (e.g., Drury 1983), so that a momentum distribution
becomes flattened to $f(p) \propto p^{-q_s}$, if initially steeper, where
$q_s = 3r/(r -1)$, and $r$ is the shock compression ratio. In all three
models electrons in smooth flows are subject to adiabatic energy gains
and losses as a result of flow expansion or compression. 

In two of the models all the nonthermal electrons are introduced only
at the jet origin, whence they are advected with the jet plasma. 
Those electron populations enter with a power law momentum
distribution, $f(p) \propto p^{-q_j}$,  with $q_j = 4.4$,
which corresponds to a
synchrotron spectral index $\alpha = 0.5(q-3) = 0.7$, typical for jets.
These two models differ only in the rates of radiative cooling for
the nonthermal electrons. In our ``control model'', radiative cooling is
negligible, whereas in the ``strong-cooling model'' an electron with
momentum $\hat p = 10^4$mc (E = 5 GeV) would loose half its
initial energy to synchrotron radiation in a field $B_{p0}$ over the full duration of the
simulation. For the parameters used to create synthetic observations
(Tregillis et al. 2001a) we used $B_{p0} = 5.8\mu$G. In that
field such electrons would radiate synchrotron emission near 2.4 GHz
and their radiative lifetimes would be about 50 million years, thus
defining the duration of the simulation. We note that since the
equations of ideal MHD have no inherent time or length scales, 
the dynamics can be computed without any reference to physical
units. Defining radiative cooling rates constrains physical
parameters, but, as discussed in Jones et al. 1999, it is still possible
to rescale the physical system in order to 
keep the dynamics unchanged while adjusting the electron cooling rates.
That is what we have done here. The intent is to explore how electron
``aging'' takes place in the simulated flows, taking into account the
highly intermittent character of the magnetic fields.

The third electron transport model injects new nonthermal electrons
from the thermal plasma at shocks and then follows them in a manner 
analogous to the control model. Evidence from galactic shocks, such
as supernova remnants, strongly suggests that a fraction of the
thermal electron population passing through a collisionless shock
is injected into the nonthermal population, although the physics of
that process is still unclear (e.g., Jones 2001). Here we apply a simple injection
model that extracts a small, fixed fraction ($10^{-4}$) of the total
electron flux through a shock. The intent
of the ``injection model'' simulation is to examine the effects of the complex 
shock web on particle
spectra when the population is dominated by in situ injection. Thus, 
radiative cooling is made negligible and the nonthermal electron population
entering at the jet origin is small, as well.

The electron spectral distributions of all three transport models are
intricate, reflecting the complex shock and magnetic field properties
discussed earlier. The chaotic motions in the cocoon 
make it difficult to identify the immediate causes of any
particular feature in a local electron population. Generally
speaking, the local particle spectra usually depend more directly on
where the particles have been than what their current local environment
is like. In a complex, changing flow, that is difficult to
reconstruct that history from snapshots.

We offer a few comments on the individual model behaviors, but for
details refer readers to a more complete discussion in 
preparation (Tregillis, Jones \& Ryu 2001b).
For the control model
the entire electron population has spectra flatter than $q_j = 4.4$,
since diffusive shock acceleration can only flatten the spectrum.
>From standard relations one can easily compute the minimum shock strength necessary to
flatten an incident electron population; namely,
$M = \sqrt{q_s/(q_s - 4)}$ (with $\gamma = \frac{5}{3}$). 
Shock modification of the electron population entering with the
jet requires passage through shocks of modest strength; namely,
$M > 3.3$.
However, just as in the earlier axisymmetric results, there is little evidence
of particle acceleration directly associated with a dominant
jet terminal shock.
Too little of the jet flow exits through such a shock structure to
have a global impact.
Instead, the evident particle acceleration is associated with strong
shock sites within the shock web near the jet terminus. Those tend
to vary quickly over both space and time.
Thus, the
particle spectral distribution shows much delicate structure, which
is difficult to capture in gray scale, but which is evident in the
color image given in Jones et al. 2000 and on the previously mentioned
web site. 

\begin{figure}
\plotfiddle{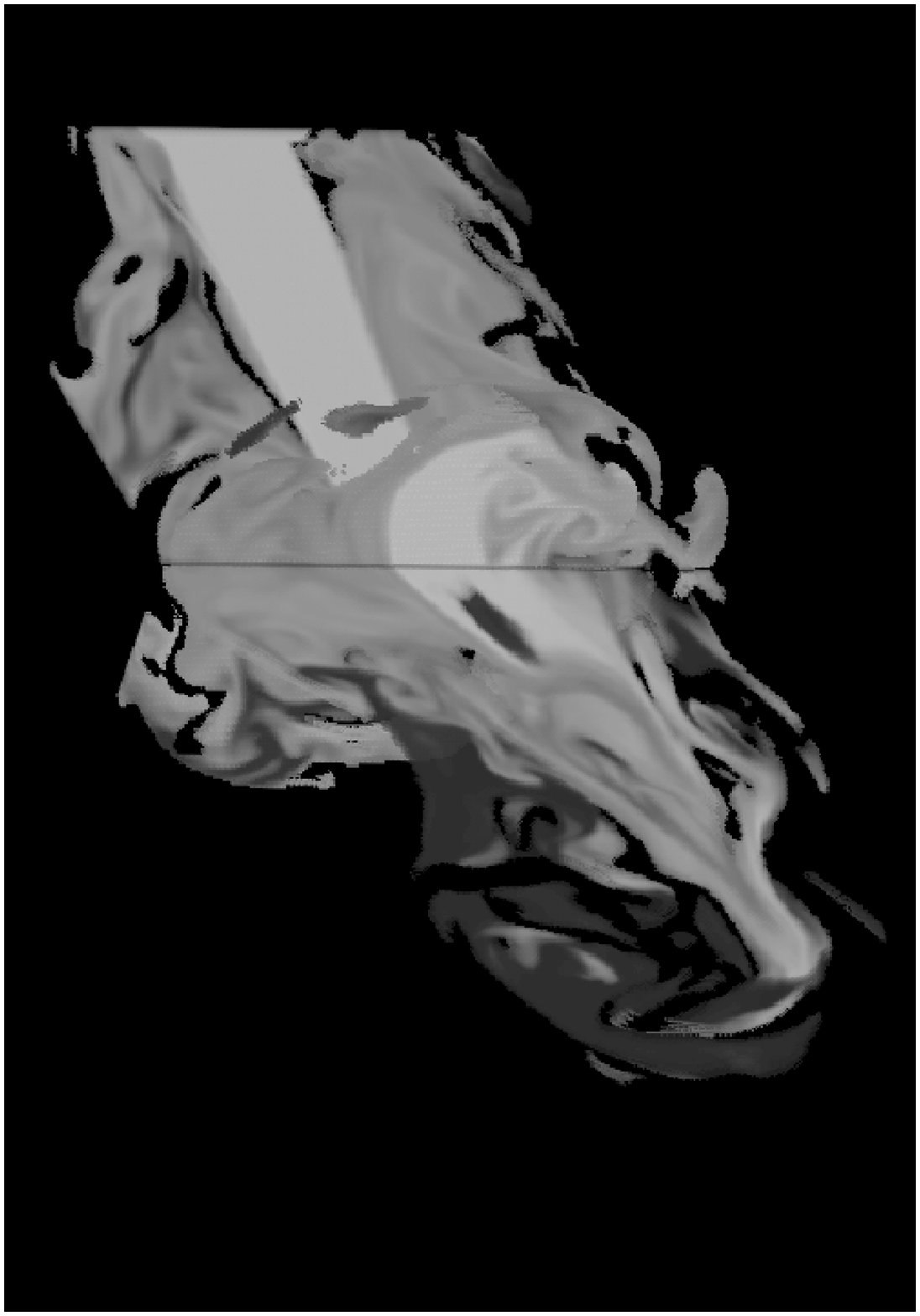}{2.25in}{-00}{30}{30}{-180}{-44}
\plotfiddle{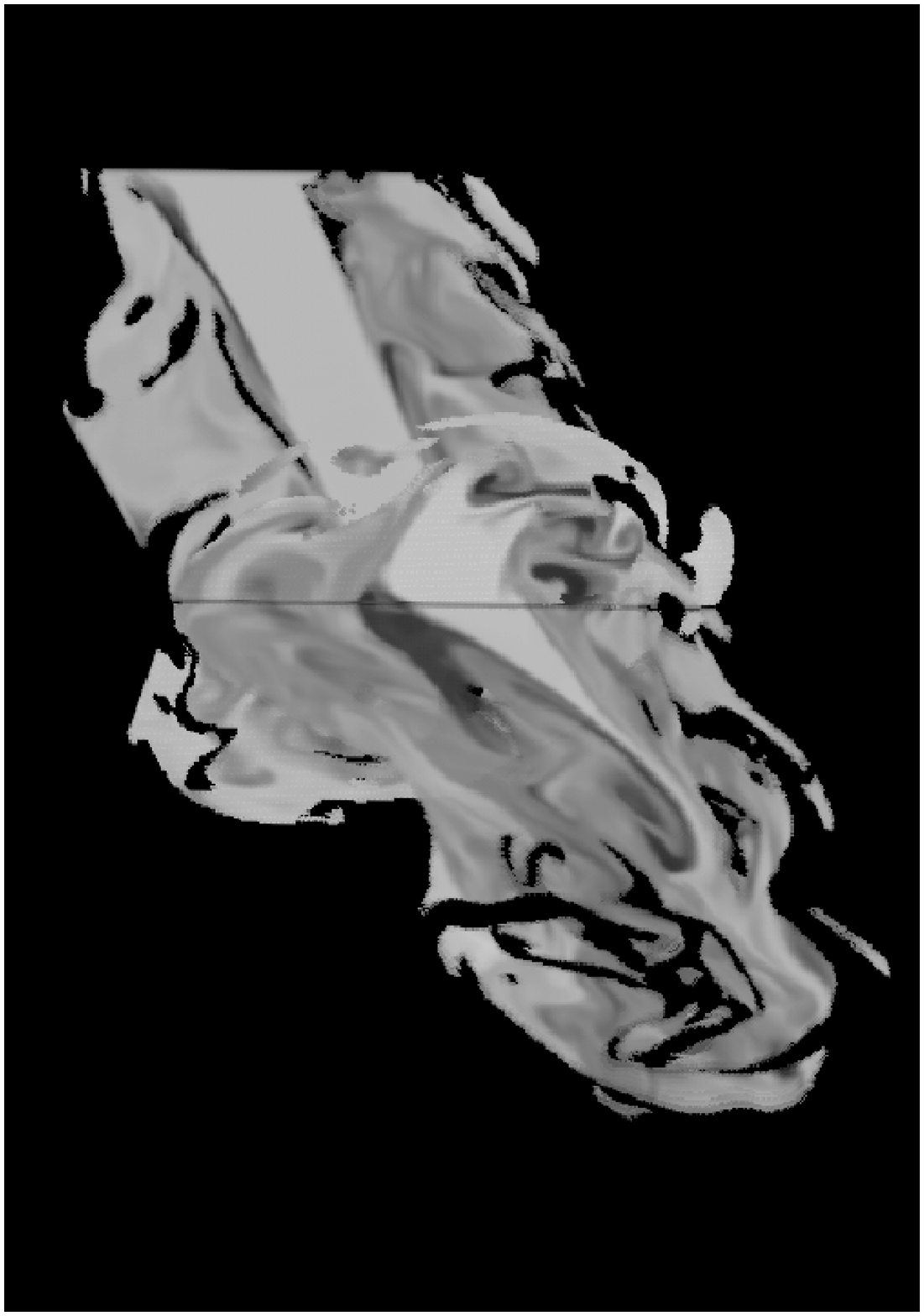}{0.0in}{-00}{30}{30}{-10}{-20}
\caption{Pairs of orthogonal slices through the grid rendering the 
spectral index of 5 GeV electrons. Left: The ``injection'' model, in 
which the electron population is dominated by fresh injection at
shocks in the flow. Radiative cooling is negligible. 
Right: The ``strong cooling'' model, in which 5 GeV electrons just cool
over the duration of the simulation if embedded in the nominal jet
magnetic field. High tones represent flatter spectra.}
\label{figure 3}
\end{figure}

The effect of the shock web is easily apparent in the electron spectral
distributions shown for the injection model in the left panel of Figure 3. 
The images in this figure come from the same time as those in
Figures 1 and 2, and the orientation of the grid is the same, as well.
They render the spatial distribution of $q$ for 5 GeV electrons,
with flatter spectra having higher tones.
The image is not weighted by emissivity, so cannot alone tell us the
expected synchrotron spectrum along a given line of sight. That property
requires a radiative transfer calculation based on self-consistent
emissivities, which is done in the companion paper by Tregillis et al. 2001a.
To view $q(5GeV)$ we have taken a 2D slice down the center of the computational
box and a transverse one as well. Again only
plasma that is entirely of jet origin is rendered. The remaining space
is transparent. As an aside we note that the irregular
boundary of the visible regions emphasizes the fact that some large
scale mechanical mixing is taking place between the jet cocoon and the CM
in response to Kelvin-Helmholtz instabilities.

The jet can be seen entering from the top.
A very small population of nonthermal electrons is actually introduced by the
in-flowing jet with $q = q_j = 4.4$ in this model, making the jet visible here. 
Almost everywhere else, the electron population
is completely dominated by injection at the shocks within the flow, however.
Mostly the electron spectral slope lies in the range $4.4 < q < 5.5$,
but there are pockets of larger and smaller values, as well.
We see for these slices that most of the volume is occupied by
electron populations with spectra moderately
steeper than $q_j$ for the incoming jet. That reflects the fact that most
of the plasma gets processed only by relatively weak shocks, as we
discussed earlier. Sometimes a particular slope value streams
into the back-flow, identifying relatively long lived shock structures
and their downstream flow patterns. This is about the closest the
patterns come to the canonical model in which all the cocoon flow represents
flow after passage through a unique, strong jet termination shock. 
The more realistic situation is clearly much more complicated.

The right panel in Figure 3 shows the analogous
electron spectral index distribution in the ``strong cooling'' transport
model. Recall in this model that all the nonthermal electrons
entered with the jet, so the only role of shocks is to flatten
portions of the momentum spectrum steeper than $q_s$
as a population enters the shock. Thus, without radiative cooling
all rendered regions in this image would show tones (``brightness'')
at least as high as the in-flowing jet.
For this model, 5 GeV electrons would just cool over the
duration of the simulation if they sat in a magnetic
field equal to the jet poloidal value, $B_{p0}$. 
Of course, electrons actually spend a much shorter time inside
the jet itself, since jet plasma would cross the computational grid in 
about $\frac{1}{10}$ the simulation time if it were not deflected by
the ambient medium along the way. So, there is little ``aging'' of
the electron population at this energy within the jet flow.
The most notable property of the distribution outside the jet is
its complex structure showing that spectral aging is not a smooth
function of location, and certainly not a simple
measure of distance from the jet terminus. 
In fact one can again identify ``streams'' of nearly constant $q$
that roughly correspond to the patterns of flow.
Mostly $4.4 < q < 5$, but both flatter and
steeper pockets are scattered through the volume. 
Thus, most of the cocoon electron population has
experienced some amount of radiative cooling at this energy, as 
intended for this model. 
On the other hand that cooling has mostly taken place very near the
head of the flow, and it is not uniform. That pattern is quite consistent,
in fact, with the properties of the magnetic field as illustrated
in Figure 2, especially when one accounts for the episodes of
stronger field amplification mentioned in the dynamical discussion.
>From these behaviors it is clearly very risky to attempt to use 
synchrotron spectral distributions to infer local spectral aging rates.

\section{Conclusions}
We have begun an effort to study the connections between dynamics in
simulated radio galaxy plasma flows and the properties of nonthermal
electron populations carried in those flows. To do this we have
introduced a new numerical scheme for electron transport that
allows us a much more detailed look at this problem than has been possible
before. Especially when the dynamics is fully three dimensional
the flows are generally chaotic in the cocoon, and the jet itself
can flail about violently. The deflected jet can pinch itself off
and redirect itself to enhance its penetration of the ambient medium. 
These behaviors mostly eliminate the
presence of a strong jet termination shock, which is assumed 
present in all modern cartoon models of the RG phenomenon. Instead a much more
complex ``shock web'' forms near the end of the jet that leads to a far
less predictable pattern of particle acceleration. Similarly,
the magnetic fields in these flows are highly filamented, as well as 
spatially and temporally intermittent. This leads to a very
localized and complex pattern of synchrotron aging for relativistic electron
populations,
which makes it difficult to use properties of the electron
spectrum to infer the local rate of aging.

These results may appear discouraging to interpretation of
observations at first glance, but we aim through this study to 
establish a more robust physical basis for linking observed
source properties with physical model characteristics, and so to
define firmly such key properties as the power of the jets
that drive these phenomena. An important
step in that task is the use of synthetic observations of the simulated
objects. That effort is also underway as described in our companion 
paper in these proceedings (Tregillis et al. 2001a).

\acknowledgments
Work by TWJ and ILT has been supported under NSF
grants AST96-16964 and AST00-71167, as well as the University of
Minnesota Supercomputing Institute. Work by DR has been supported by
KOSEF grant KRF-2000-015-DS0046.

\end{document}